\documentclass[12pt, journal, draftcls, onecolumn]{IEEEtran}
\usepackage{algorithm}
\usepackage{algorithmic}
\usepackage{amsfonts}
\usepackage{amssymb}
\usepackage{cite}
\usepackage{color}
\usepackage{multirow}
\usepackage{subfigure}
\usepackage{graphicx,times,amsmath}

\begin{document}

\title{Toward Cross-Layer Design for Non-Orthogonal Multiple Access: A Quality-of-Experience Perspective}

\author{Wei~Wang,
	Yuanwei~Liu,
	Zhiqing~Luo,
	Tao~Jiang,~\IEEEmembership{Senior~Member,~IEEE},
	Qian~Zhang,~\IEEEmembership{Fellow,~IEEE},
	Arumugam~Nallanathan,~\IEEEmembership{Fellow,~IEEE}
	\thanks{W. Wang, Z. Luo and T. Jiang are with the School of Electronic Information and Communications, Huazhong University of Science and Technology. E-mail: \{weiwangw,zhiqing\_luo,taojiang\}@hust.edu.cn.}
	\thanks{Y. Liu is with Queen Mary University of London, UK. E-mail: yuanwei.liu@qmul.ac.uk.}
	\thanks{A. Nallanathan is with King's College London, London, UK. E-mail: arumugam.nallanathan@kcl.ac.uk.}
	\thanks{Q. Zhang is with the Department of Computer Science and Engineering, Hong Kong University of Science and Technology.  E-mail: qianzh@cse.ust.hk.}}

\maketitle

\begin{abstract}
Recent years have seen proliferation in versatile mobile devices and an upsurge in the growth of data-consuming application services. Orthogonal multiple access (OMA) technologies in today's mobile systems fall inefficient in the presence of such massive connectivity and traffic demands. In this regards, non-orthogonal multiple access (NOMA) has been advocated by the research community to embrace unprecedented requirements. Current NOMA designs have been demonstrated to largely improve conventional system performance in terms of throughput and latency, while their impact on the end users' perceived experience has not yet been comprehensively understood. We envision that quality-of-experience (QoE) awareness is a key pillar for NOMA designs to fulfill versatile user demands in the 5th generation (5G) wireless communication systems. This article systematically investigates QoE-aware NOMA designs that translate the physical-layer benefits of NOMA to the improvement of users' perceived experience in upper layers. We shed light on design principles and key challenges in realizing QoE-aware NOMA designs. With these principles and challenges in mind, we develop a general architecture with a dynamic network scheduling scheme. We provide some implications for future QoE-aware NOMA designs by conducting a case study in video streaming applications.
\end{abstract} 
\section{Introduction}
Recent years have witnessed a boom in versatile wireless devices and data-consuming mobile services. In the coming wave of a large number of new generation mobile Internet devices, including tablets, smartphones, smart wearables and so on, future 5th generation (5G) mobile networks need to support the massive connectivity of wireless devices. Additionally, the emergence of new data-consuming services, such as virtual reality (VR), augmented reality (AR), high-definition (HD) video streaming, cloud and fog computing services, have elevated the traffic demands.


The ever-increasing traffic demands have raised the stakes on developing new access technologies to utilize limited spectrum resources. Non-orthogonal multiple access (NOMA), as an emerging multiple access (MA) technology for improving spectral efficiency, has recently obtained remarkable attention~\cite{dai2015non,boccardi2014five}. The innovative concept of NOMA is to serve multiple users in a single resource block, and thus is fundamentally different from conventional orthogonal multiple access (OMA) technologies, such as time division multiple access (TDMA) and orthogonal frequency division multiple access (OFDMA). Recently, a downlink version of NOMA, termed multiuser superposition transmission (MUST), has been included for the 3rd generation partnership project (3GPP) long term evolution (LTE) initiative~\cite{LTE2015}. Another NOMA technology, namely layer-division-multiplexing (LDM) has been accepted by the digital TV standard advanced television systems committee (ATSC) 3.0~\cite{zhang2016layered} for its efficiency in delivering multiple services in one TV channel.

Despite growing attempts and extensive efforts on NOMA, most studies have focused on the physical layer (PHY) or medium access control layer (MAC) performance, such as throughput and PHY latency, while few have systematically investigated its impact on user's perceived quality of experience (QoE). QoE is the perceptual quality of service (QoS) from the user's perspective~\cite{qoesurvey}. Given the context that the growth of future traffic demands is largely driven by the visual-experience-oriented services, such as VR, AR, and video streaming, there is broad consensus among leading industry and academic initiatives that improving user's QoE is a key pillar to sustaining the revenue of service providers in future networks~\cite{qoe}. {Despite this agreement on the importance of QoE, our understanding of how NOMA affects QoE is limited. The reason is that NOMA is a PHY/MAC access technology whose primary goal is to improve spectrum efficiency in the presence of massive connectivity, while QoE is an upper-layer concept related to user engagement and end-to-end quality.}

The goal of this article is to explore the upper-layer impact of NOMA on the user side, and call attention to a clean-slate redesign of cross-layer NOMA frameworks for QoE provisioning. {The remainder of this article is structured as follows. In Section II, we start at a deep dive into NOMA system architectures, and then review QoE demands and the challenges of realizing QoE awareness in NOMA systems. Specifically, we shed light on QoE metrics and confounding factors in NOMA systems, as well as how the lower-layer NOMA strategies affect user experience in the upper layers. Then, we follow on the heels of implications from the QoE analyses in NOMA, and propose a QoE-aware framework for NOMA systems in Section III. We build a model based on real-world datasets from a cellular service provider and take a case study on video streaming applications in Section IV. Merits of the QoE-awareness framework are verified, and implications about future QoE-aware NOMA schemes are provided in Section V.}
\section{Exploiting QoE Awareness in NOMA}
In this section, we describe the system architecture of NOMA from a cross-layer perspective, and discuss user experience issues that arise with NOMA. In particular, we highlight QoE demands and confounding factors in NOMA systems. We also investigate how the user experience in the upper layers is affected by NOMA in PHY/MAC.

\subsection{NOMA Premier}

\begin{figure}[t]
	\centering
	\includegraphics[width=5.5in]{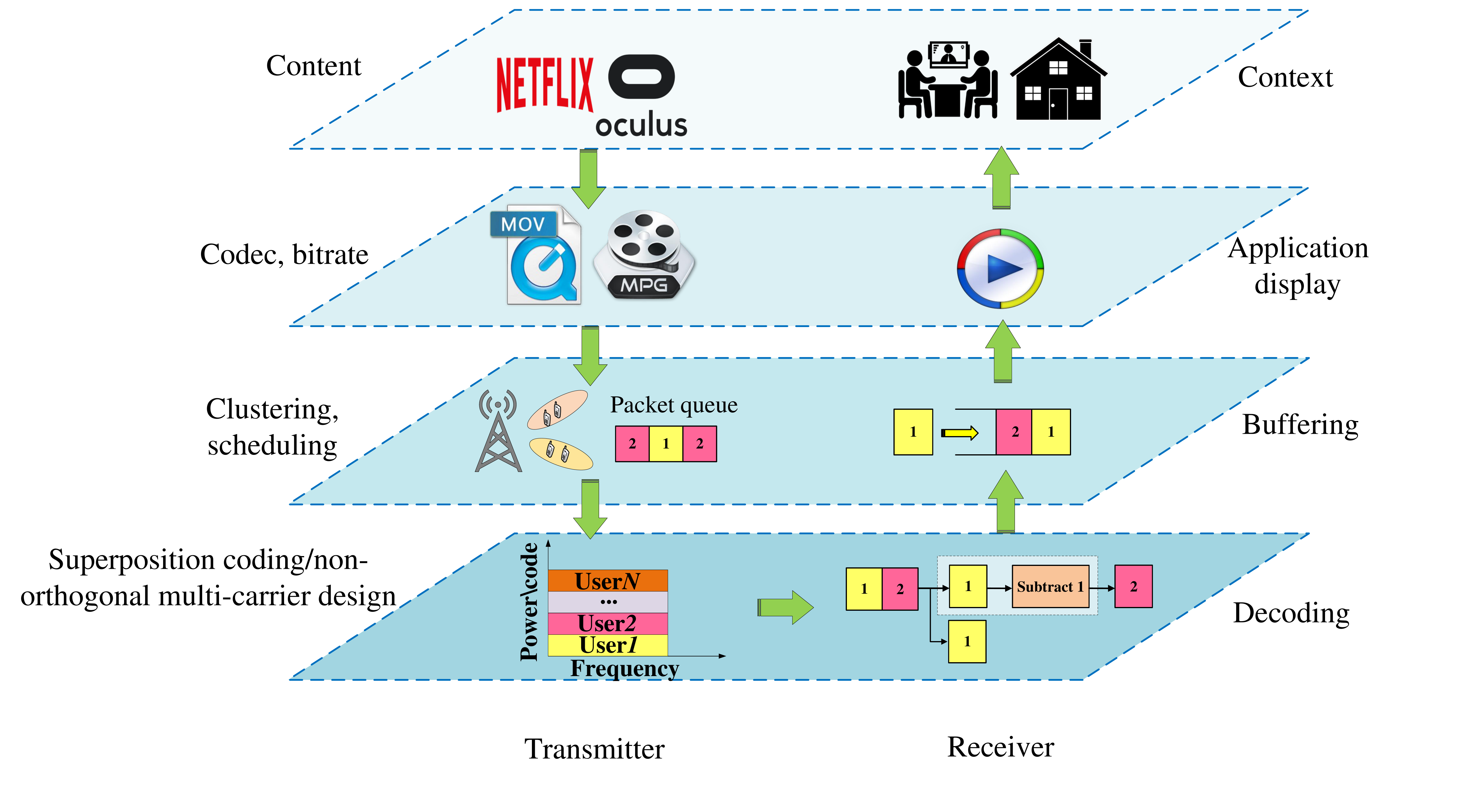}
	\caption{NOMA systems: a cross-layer perspective.}
	\label{fig:system}
\end{figure}
The essential idea of NOMA is allocating non-orthogonal resources to multiple users by allowing controllable interference and tolerable decoding complexity at receivers. Basically, the gamut of NOMA technologies is divided into a pair of categories, that is, \emph{power-domain NOMA} and \emph{code-domain NOMA}. Power-domain NOMA multiplexes several users in the same subcarrier, by employing superposition coding (SC) technology at transmitters. At receivers, successive interference cancelation (SIC) technology is applied as such multiple users can be distinguished via different power levels~\cite{ding2015application}. Code-domain NOMA departs from power-domain NOMA in that it adopts multi-carrier operations. More particularly, it mainly relies on applying low density or non-orthogonal sequence designs at transmitters over multiple carriers, then invoking joint detection technology such as message passing algorithms (MPA) at receivers for obtaining coding gains. Representative code-domain NOMA technologies include sparse coding multiple access (SCMA)~\cite{nikopour2013sparse}, patten division multiple access (PDMA), and low density signature code domain multiple access (LDS-CDMA)~\cite{hoshyar2008novel}. 

Fig.~\ref{fig:system} illustrates the system architecture of downlink NOMA from a cross-layer view. At the top level, content is delivered to users, who are in different contexts such as at home, in office shops, and in transportation. To this end, the content is first encoded into bit streams that are passed from the application layer to the lower layers, where packets for different users are scheduled and transmitted in a non-orthogonal manner. In particular, the power-domain multi-carrier NOMA first partitions users into different clusters, and assigns each cluster to one orthogonal subcarrier~\cite{ali2016dynamic}. Normally, effective user scheduling algorithms and power allocation approaches are employed. Regarding code-domain NOMA that is originally multi-carrier based, appropriate codeword selection and power level adjustment at each subcarrier need joint consideration. Receivers perform corresponding decoding algorithms to extract their own information from the received frames at PHY. In visual-based applications, received data is buffered or played according to the data rate of the downlink, video quality, and display speed. Agnostic to all the above data transmissions and processes, users only sense and feel the services, basically the content delivered to and displayed on their own devices. We observe from Fig.~\ref{fig:system} that NOMA functionalities including user clustering, packet scheduling, and power allocation, are the cornerstones of the upper-layer service quality delivered to the end users. In the following sections, we investigate the user experience in the NOMA system, and take one step further to analyze the impact of NOMA functionalities on the user experience. {Architecturally, such a cross-layer design can be realized by following on the heels of the software-defined network (SDN) paradigm. In particular, a controller extracts the QoE requirements of different types of service from the upper layers. All the collected information is forwarded to lower layers to guide traffic scheduling and resource allocation in NOMA.}
\begin{figure}[t]
	\centering
	\includegraphics[width=4.5in]{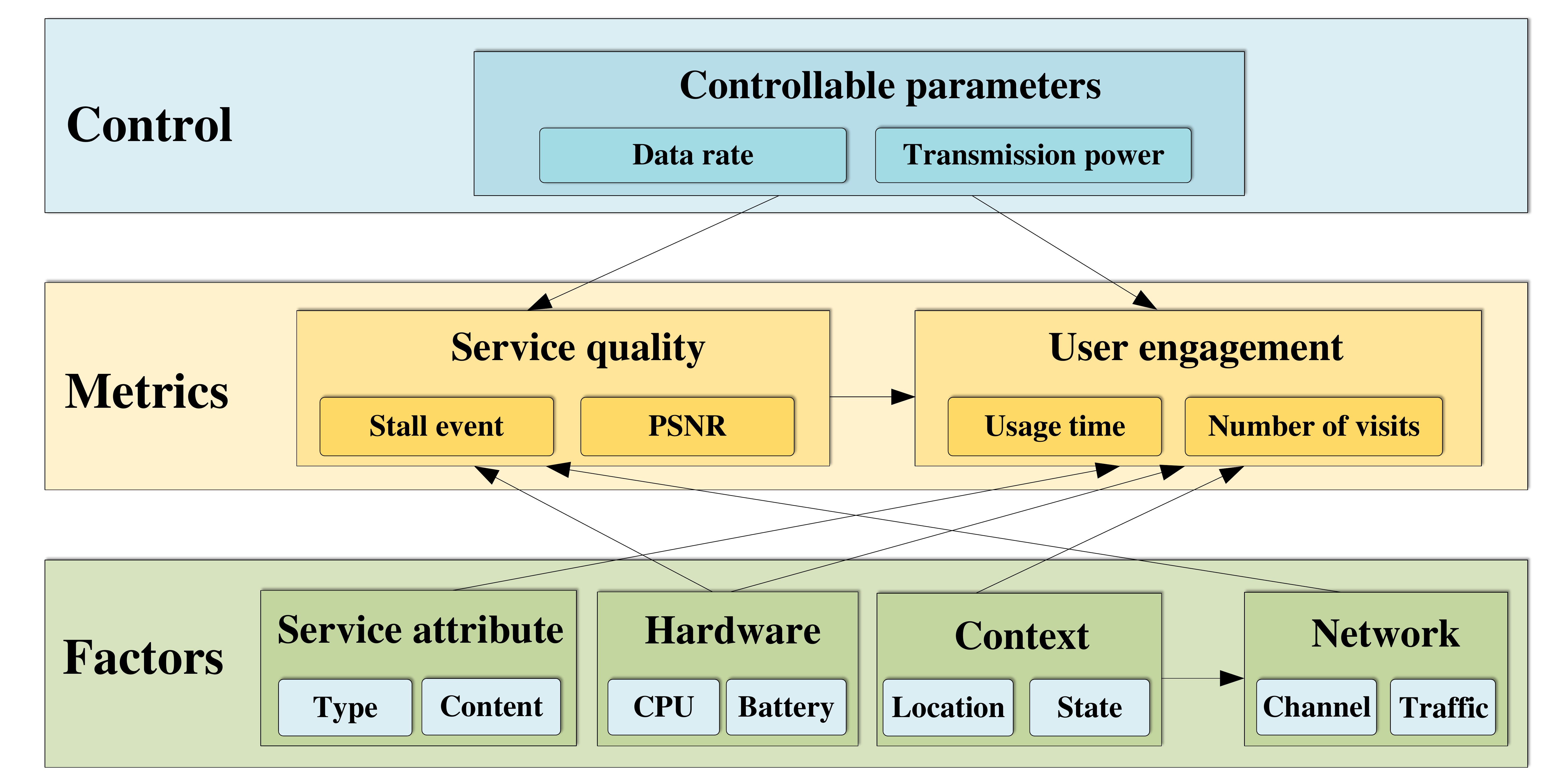}
	\caption{Complex interdependencies between QoE metrics and confounding factors. Controllable parameters at PHY also have impact on service quality or user engagement.}
	\label{fig:metric}
\end{figure}

\subsection{QoE Metrics and Confounding Factors in NOMA Systems}

NOMA is conventionally evaluated using PHY connection-centric metrics such as throughput or sum-rate capacity. However, conventional connection-centric designs have become a barrier to meeting the diverse application requirements and the quality expectation of the end users, especially for the rapid expanding visual-experience-oriented services, such as VR, AR, and video streaming. Even though the data rate and throughput are increasing, current mobile networks are still facing poor user experience and low service quality~\cite{qoe}. The gap between the PHY system performance and higher-level user's perceived experience drives a paradigm shift from connection-centric designs to experience-centric designs. QoE is a metric that quantifies user's perceived experience from a subjective perspective~\cite{qoesurvey}. To fully understand NOMA from the QoE perspective, we first investigate the complex relationship between the QoE metric and confounding factors in NOMA that may affect the user experience. Interdependencies between metrics and factors are depicted in Fig.~\ref{fig:metric}.

Various metrics can be used to quantify user's different subjective experience. They are mainly divided into engagement metrics and quality metrics. Engagement metrics such as \textit{usage time} and \textit{number of visits} reflect user's satisfaction and content's popularity. Given a certain context and service content, user engagement largely depends on the service quality. The commonly-used industry-standard quality metrics include \textit{peak signal-to-noise ratio (PSNR)} and latency-wise metrics such as \textit{join time} and \textit{stall event}. PSNR is most commonly used to measure the quality of reconstruction of lossy compression codecs in visual-based services. Join time represents the time it takes for the service to start offering content (such as video) after the user initiates a request, which indicates the start-up delay and thereby leading to different levels of satisfaction for the service. The number of stall events is a crucial indicator of end-to-end latency during the service usage and undermine user experience. 

\textbf{Confounding factors}. There are many confounding factors impacting on user engagement and service quality. Confounding factors mainly include network conditions, hardware capabilities, context and service attributes.
\begin{itemize}
	\item \textbf{Network conditions} include traffic, channel, and interference conditions. Since bad network condition will significantly degrade user experience, many QoE models regard network condition as one of the most influential factors for user's QoE. 
	\item \textbf{Hardware capabilities} including CPU processing ability, battery condition, and screen size affect user's QoE demands as well as perceived experience. For example, the processing capability of CPU largely determines player's start-up delay and users of low-end devices are more concerned with the CPU resources required by the service.
	\item \textbf{Context} includes location, user's state (such as driving, walking, dining), and surrounding environments (such as office, coffee shop). Context also indirectly affects service quality as different contexts may result in different network conditions (such as interference and channel conditions).
	\item \textbf{Service attributes} include type, content and popularity. In general, users tend to spend more time using services with their preferred attributes, which are valuable references for QoE evaluation.
\end{itemize}

\subsection{Understanding NOMA from the Perspective of QoE Awareness}
Conventionally, NOMA is proposed as a novel PHY/MAC multiple access paradigm and is designed to support massive connectivity with low latency and diverse service types. To fully reap the benefits of NOMA to improve end-to-end performance and user's perceived experience, we investigate how NOMA in PHY/MAC affects user experience in the upper layers.


\subsubsection{QoE Awareness in Transmission Pipeline}
The crux of NOMA is to assign multiple users to the same time/frequency/code resource units with tolerable interference. From the PHY-level perspective, two or more users are clustered to achieve maximal sum-rate capacity or throughput with fairness or latency constraints~\cite{ding2016impact}. However, from the user's perspective, it does not provide QoE guarantee for each user. Recall that QoE demands are heterogeneous due to user's preference, hardware diversity, context difference and so on, and thus data rate or throughput cannot ensure user's QoE. To overcome this predicament, it is required to translate the upper-layer demands and diversity into PHY objectives and constraints. The confounding factors such as service types and hardware conditions determining QoE metrics should be considered when clustering users in NOMA. Users with low-end devices or high sensitivity for latency can be clustered with delay-tolerant users with residual computational power. 


Power allocation and packet scheduling schemes in NOMA should also be tailored to support QoE provisioning. Conventionally, {power-domain NOMA allocates different users with different power levels for achieving a good throughput-fairness tradeoff~\cite{ding2015application}. Regarding code-domain NOMA, it relies on sparse or low density spreading sequences to multiplex users on multiple subcarriers for realizing overloading, with the aid of the so-call coding gain~\cite{nikopour2013sparse}.} From the QoE perspective, power allocation schemes should reap the benefits of NOMA without compromising individual's QoE. Likewise, packet scheduling schemes in NOMA need to be reformulated to support QoE awareness. User engagement and quality metrics should be considered as constraints or objectives, and factors such as hardware, service type, and context information should be taken into account.


\subsubsection{QoE Awareness in Reception Pipeline}
At the receiver, a decoding algorithm with increased complexity is normally employed to cancel multi-user interference in NOMA. The extra cost in decoding NOMA packets may not be affordable to devices with limited remaining battery or computational capabilities, and thus may shorten usage time or cause CPU process delay that undermines user's QoE. 

The decoding issue is more complex in power-domain NOMA, in which SIC is performed at receivers in a descending order of signal-to-interference-plus-noise ratio (SINR). Once a receiver fails to decode a packet in SIC, the following packets become undecodable. On the one hand, multiplexing more users in the same resource unit can potentially achieve higher sum rates, thereby amortizing network congestion. With proper configurations and channel conditions, multiplexing more users creates greater potential to satisfy more users QoE demands. On the other hand, it increases the probability of decoding failure and thus undermines the robustness of packet transmission. This issue becomes more severe in dynamic, unpredictable radio environments such as buses or trains. The sacrifice of robustness may lead to unstable service quality such as stall events in video streaming. Additionally, it incurs substantially higher computational overhead at users with low SINR, which drains battery and may incur processing delay for low-end devices. As a result, NOMA schemes are expected to consider the decoding capabilities at user end and the corresponding QoE demand to play proper strategies.








\begin{figure}[t]
	\centering
	\includegraphics[width=6in]{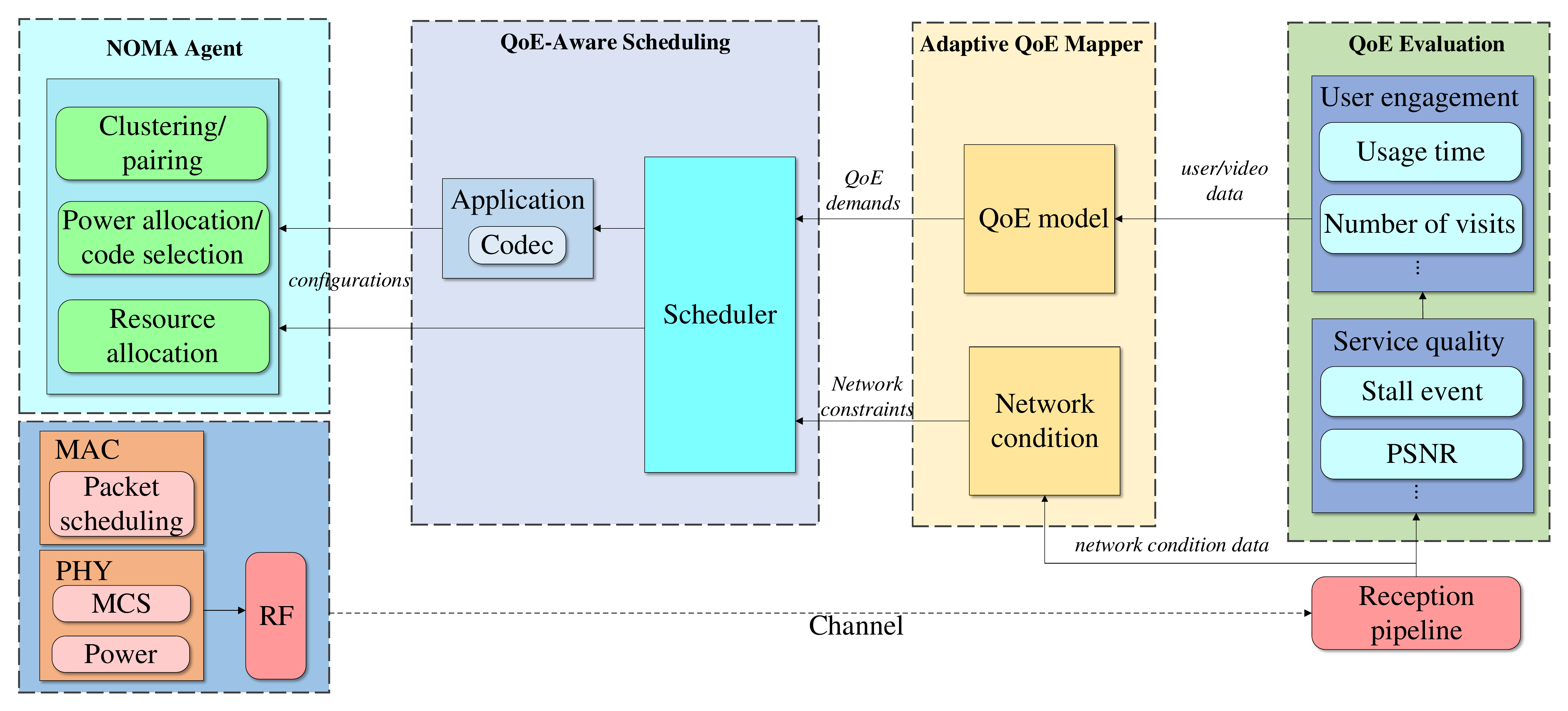}
	\caption{QoE-aware NOMA framework.}
	\label{fig:framework}
\end{figure}

\section{QoE-Aware NOMA Framework}

In this section, we develop a cross-layer NOMA framework that provides QoE guarantee for the NOMA system.

\subsection{Overview}
Fig.~\ref{fig:framework} outlines the proposed QoE-aware framework that facilitates the NOMA system to schedule users and packets in order to fit user's diverse QoE demands. In particular, the key design components are described in a top-down fashion.
\begin{itemize}
	\item \textbf{QoE Evaluation}. The function of QoE evaluation is to collect user data and estimate QoE demand for each user. User data consists of two parts: historical data and real-time data. The historical data contains complete datasets of user engagement, service quality, confounding factors and corresponding NOMA configurations and parameters. These datasets are used to train a QoE demand profile for each user indicating their preferences. 
	\item \textbf{Adaptive QoE mapper}. The adaptive QoE mapper translates QoE demands from users to objectives and requirements described by controllable system parameters in NOMA. The mapper is built based on the QoE model trained in QoE evaluation and the real-time network condition feedback. It interacts with the network protocol stack through a scheduler.
	\item \textbf{Scheduler}. The scheduler acts as a control plane that derives feasible or optimal configurations to fulfill the requirements given by the adaptive QoE mapper. The configurations, {including power allocation, user clustering, queuing, and resource allocation strategies,} are tuned based on the updated QoE model as well as network condition feedback from users or measured by the transmitter. 
	\item \textbf{NOMA agent}. The NOMA agent resides atop PHY/MAC and acts as an interface between NOMA functionalities and the scheduler. It exposes commands from the scheduler to NOMA functionalities to perform user clustering/pairing, power allocation, and packet scheduling.
\end{itemize}

In the following two sections, we shed light on how to implement the above framework to achieve QoE guarantee for different users in NOMA with adaptive scheduling.

\subsection{Adaptive Mapping for QoE-Aware NOMA}
The design rationale of QoE-aware NOMA is in a top-down fashion: we start with user's perceived QoE and map QoE demands into system parameters that are taken as input to derive proper NOMA configurations. Since QoE can be quantified from different perspectives, we perform a data-driven QoE analysis based on  user data collected by service providers. For each user, we identify a subset of factors that have the most significant impact on QoE while screening out insignificant factors to reduce dimension and noise. {To this end, we measure the importance of a factor using information gain, which represents the informative level of a feature. Then, we rank all factors based on information gain and select the top-$k$ factors.}


Since QoE greatly depends on the user's personal preference as well as the context, the QoE of different users can be quite disparate even for the same content and service quality. To take into account user disparity, we build a distinct QoE model for each user based on collective matrix factorization~\cite{singh2008relational}, which is used to discover the latent features underlying the interactions between user engagement and service quality. The users' QoE is represented by a matrix where each entry indicates the QoE of a user for a service. Similarly, user preferences, contexts and attributes such as age, gender, and data plan form a user matrix; and a service matrix describes service attributes such as bitrates, codecs and other system parameters in NOMA. Our key idea is to jointly factorize QoE matrix, user matrix, and service matrix into two low-rank matrices describing latent factors in NOMA. The underlying intuition is that, by transforming both user and service matrices to the same latent factor space, we can estimate the user preference to these latent factors and content score on these factors. We iteratively use all the data of services used by a user to train the user's preference vector, and feed all the data of users who have used the service into a model to learn the video score vector. Finally, the factorization results are combined in a weighted sum to establish a personalized QoE model that maps QoE demands to controllable parameters in the NOMA system.

\subsection{QoE-Aware Scheduling in NOMA}
The adaptive QoE mapper translates users' QoE demands into system parameters and requirements that are controllable to the NOMA system. The scheduler takes these parameters and requirements as input to derive proper scheduling and configuration strategies for NOMA.

To sustain users' QoE demands in a long-term way, the scheduling strategies are derived by formulating the problem as a dynamic scheduling problem. We consider a scenario where the time of service $T$ is descretized into slots $t$. Our goal is to obtain an optimal scheme to cluster users in NOMA and schedule resources to minimize the average QoE loss for all the time slots, while satisfying the QoE requirements for all the users in NOMA. For example, in a video streaming case, we minimize time-averaged total quality loss for all video streams while maintaining the long-term fluency by controlling buffer data to restrain the number of stall events. In each time slot, the scheduler checks the data queue and network conditions. Then, it derives the optimal scheduling strategy by making a joint decision to reallocate resources and update the queue. 
\begin{figure}[t]
	\centering
	\includegraphics[width=5.5in]{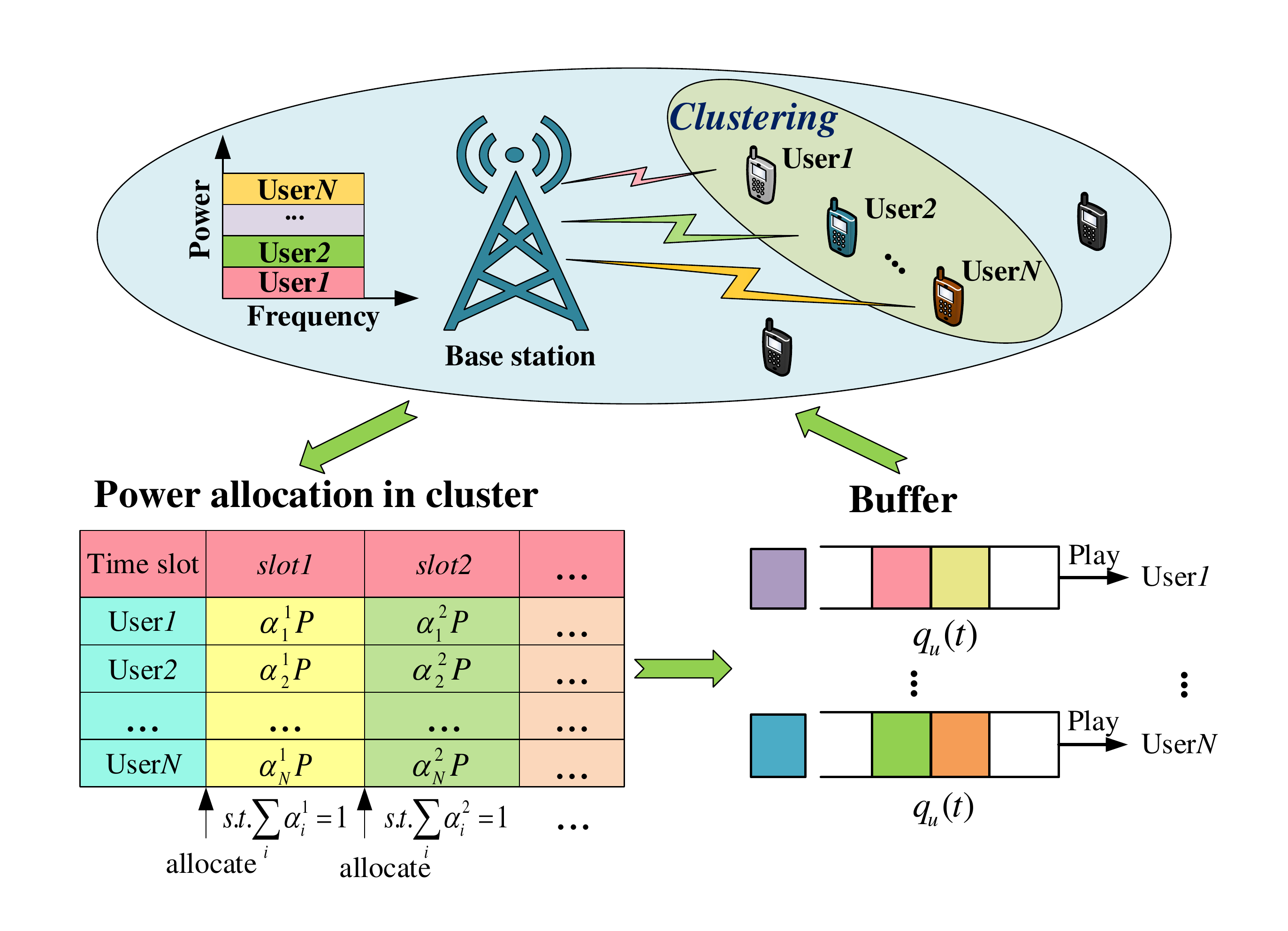}
	\caption{QoE-aware scheduling in NOMA.}
	\label{fig:schedule}
\end{figure}
In order to manage this problem, as shown in Fig.~\ref{fig:schedule}, the user clustering strategies need to be determined. Recall that confounding factors such as context, network conditions, and hardware capabilities largely determine the performance of user clustering from the QoE perspective. We set constraints for user clustering strategies based on QoE requirements imposed by these confounding factors, and maximize $\sum_u q_u(t) \cdot p_u(t) - \omega \sum_u Q_u(t)$, where $q_u(t)$ is the data queue for user $u$ in slot $t$, $p_u(t)$ the {playing duration (the duration of the video been played by the user)} of data received by user $u$ in slot $t$, $Q_u(t)$ the QoE loss of user $u$ in slot $t$ and $\omega$ the control parameter of the drift plus penalty obtained from users' QoE preferences. In addition, we allocate resource units and power to clustered users. For illustration, we only focus on the power-domain NOMA. In each slot, users are allocated with different proportions of the total power $P$ in the same frequency/time/code domain so that users can perform SIC for packet decoding. Note that transmission power and channel conditions determine the data rates of clustered users in NOMA, which decides $p_u(t)$ along with the bitrate in the application layer. We select the optimal power allocation and scheduling scheme that delivers services with the minimal QoE loss on average while providing QoE guarantee for each user.

\begin{figure}[t]
	\centering
	\includegraphics[width=5in]{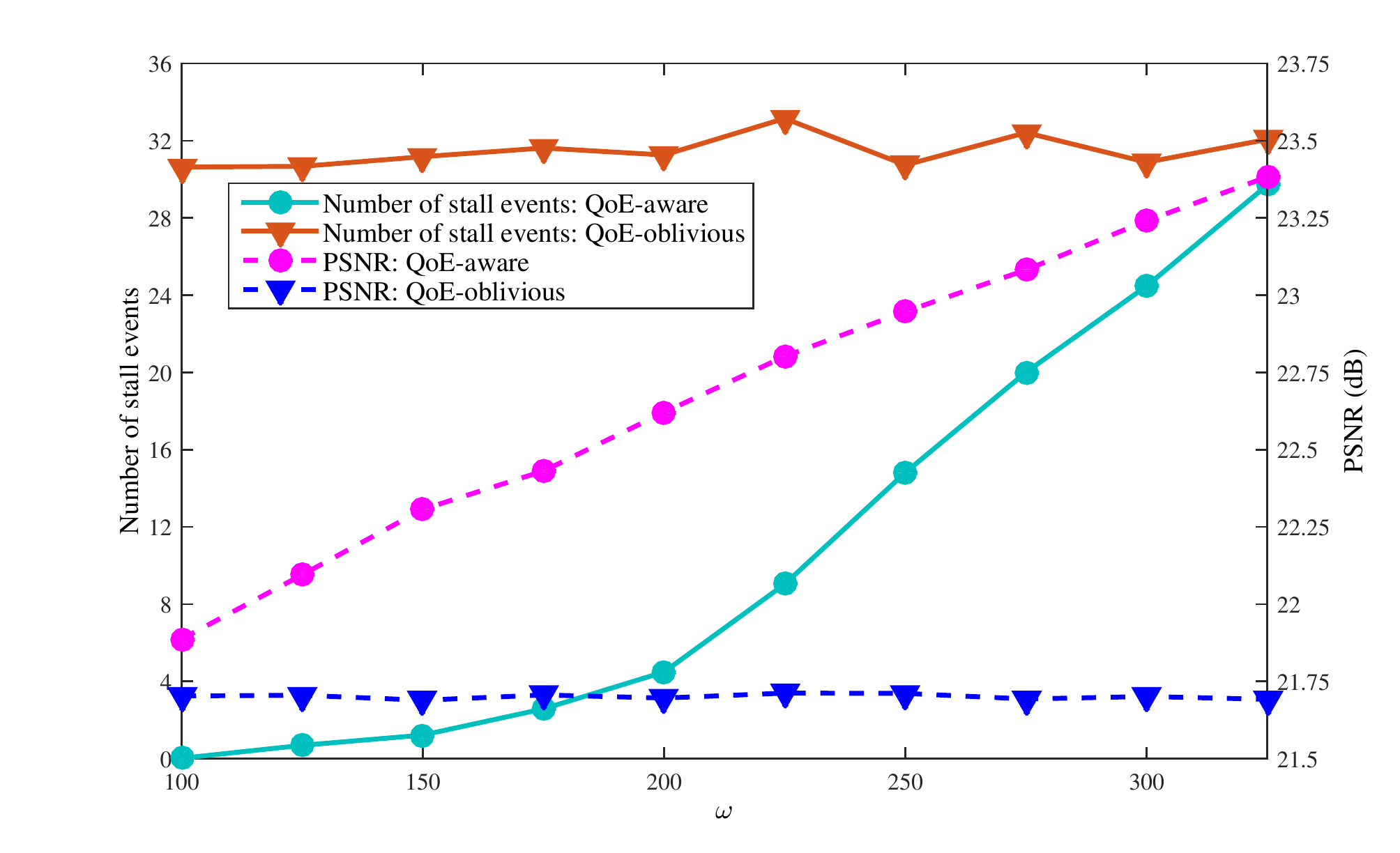}
	\caption{Performance under various values of $\omega$ in NOMA.}
	\label{fig:omega}
\end{figure}

\section{Case Study on Video Streaming}
In this section, we evaluate the QoE-aware NOMA framework using data-driven simulations. We consider power-domain NOMA to illustrate the merits. To model QoE requirements, we collected a large-scale dataset from a tier-one cellular service provider in China containing traces of eight million users over eight months, including the complete raw IP flow traces in the core network and user profile database. We build a reliable QoE model for each user to construct the QoE evaluation and adaptive QoE mapper components in the framework. To evaluate our scheduling scheme, we turn to simulations. {We consider a scenario where four users request video streaming services from one base station, whose transmission power is 20~dBm. Users are randomly distributed and move randomly within the radius of the base station. Rayleigh fading is used to model wireless channel. We use the Sony Demo, which is encoded in MPEG4 with four different quality levels, as the video streaming data. Video streaming services in which the video data is divided into sequential chunks encoded in different bitrates and quality levels. In each slot, video chunks are encoded with different quality levels and buffered at the user side to sustain video playing.}

We compare our framework with the conventional framework where QoE demands are oblivious. {In the QoE-oblivious NOMA system, resource allocation follows the max-sum throughput rule, that is, the power allocation and channel access mainly depend on the network conditions. Differently, QoE-aware NOMA considers users' QoE demands by minimizing time-averaged total quality loss while maintaining the long-term fluency by controlling buffer data to restrain the number of stall events.} We evaluate the performance under various values of control parameter $\omega$, which indicates the importance of service quality valuated by users. {Note that the QoE-oblivious scheme does not depend on $\omega$ and is used as a reference to assess the merits of the QoE awareness.} 

\begin{figure}[t]
	\centering
	\includegraphics[width=5in]{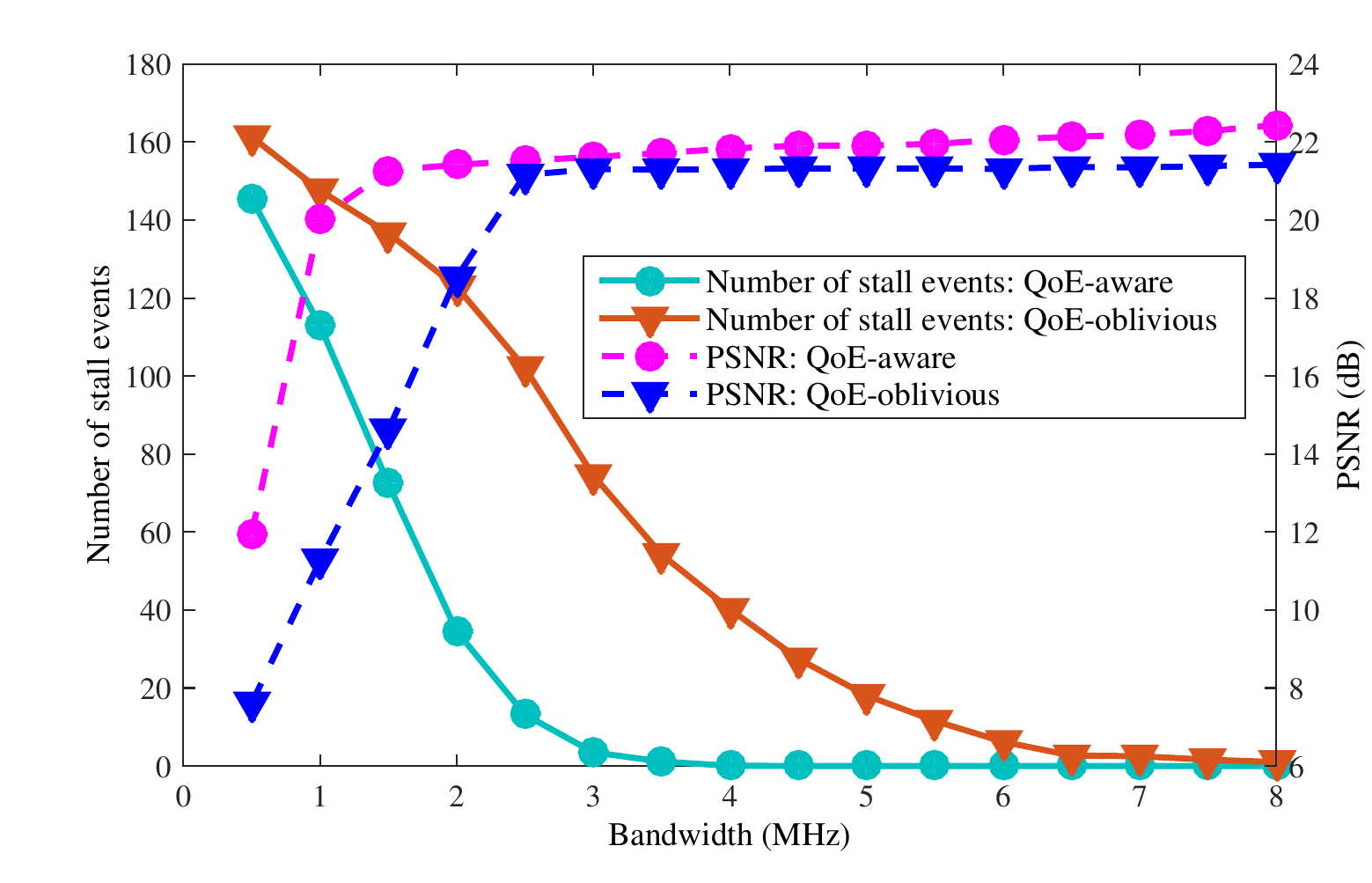}
	\caption{Performance under various bandwidths in NOMA.}
	\label{fig:bandwidth}
\end{figure}
In this study, we first evaluate the performance under various values of the control parameter $\omega$, which indicates the importance of service quality valuated by users. As shown in Fig.~\ref{fig:omega}, the PSNR of QoE-aware NOMA is higher than that of QoE-oblivious NOMA, and the performance gain of QoE-aware NOMA increases with the value of $\omega$. It reveals that QoE-aware NOMA can better fit the diverse demands of users in that it provides better service quality for users with higher quality preferences. 

The performance of the framework is also influenced by the network condition, such as available bandwidth. As illustrated in Fig.~\ref{fig:bandwidth}, both schemes achieve higher PSNR and fewer stall events with larger bandwidth. QoE-aware NOMA outperforms QoE-oblivious NOMA in all cases demonstrated, which indicates that QoE awareness can better adapt to the changes of bandwidth.

\section{Concluding Remarks and Future Directions} 
This article has envisioned the crucial role of QoE awareness in NOMA systems. Instead of merely focusing on the PHY/MAC system performance, QoE awareness exploits interdependencies among NOMA functionalities, confounding factors and user's perceived experience. Through careful investigation of the interplay between upper-layer demands and lower-layer configurations, we have presented a QoE-aware NOMA framework and demonstrated its merits by conducting a video streaming case study using real-world traces and simulations. Some observations in the case study can provide some implications for future designs of QoE provisions in NOMA.

However, the study of QoE-aware NOMA is still in its infancy. There are many open problems that need further investigation.
\begin{itemize}
	\item \textbf{Complexity-aware NOMA}. NOMA improves spectrum efficiency at the cost of increased decoding complexity at the receiver side~\cite{dai2015non}. For example, power-domain NOMA requires receivers to perform SIC for decoding. The extra computations induced by NOMA add burdens at the user end and lead to extra processing delay. Such impacts are not considered in conventional NOMA designs, while they may undermine user's QoE. How to quantify these impacts and design complexity-aware NOMA is a new direction in designing QoE-aware NOMA.
	\item \textbf{Energy-aware NOMA}. A typical application scenario of NOMA is 5G smart devices such as the Internet of things (IoT), wearables and smartphones with massive connectivity~\cite{ding2016mimo,wang2017spectrum}. These devices, however, are power-constrained. Users may care more about their energy consumption and consider battery life as a dimension of QoE. However, NOMA introduces extra computation and even communication overhead that consumes the user's battery. There will be new trade-offs in NOMA when energy consumption is considered in QoE. How to jointly consider energy consumption and service quality in QoE-aware NOMA is still an open problem.
	\item \textbf{QoE-aware NOMA in different applications}. It has been shown in~\cite{ding2015application} that NOMA can be applied to different communication scenarios, including device-to-device (D2D), multiple-input-multiple-output (MIMO), and cooperative transmission. This article investigates the general interdependencies between QoE and NOMA, while there are still many design challenges and practical issues in realizing QoE-aware NOMA in these applications.
	\item \textbf{Software-Defined NOMA}. {Recently the software-defined network (SDN) has been deemed as a new paradigm to improve spectrum manegement~\cite{wang2016software} from the network architecture perspective. NOMA offers new PHY opportunities to improve spectrum efficiency. NOMA and SDN can be jointly considered in future system designs to boost spectrum efficiency.}
	\end{itemize}

\section*{Acknowledgment}
The research was supported in part by the National Science Foundation of China under Grant 61502114, 91738202, 61729101, and 61531011, Major Program of National Natural Science Foundation of Hubei in China with Grant 2016CFA009, 2015ZDTD012, the RGC under Contract CERG 16212714, 16203215, ITS/143/16FP-A.


\begin{IEEEbiography}{Wei Wang (S'10-M'15)} (gswwang@connect.ust.hk) is currently a professor in School of Electronic Information and Communications, Huazhong University of Science and Technology. He received his Ph.D. degree in Department of Computer Science and Engineering from Hong Kong University of Science and Technology. He served as guest editors of Wireless Communications and Mobile Computing, IEEE COMSOC MMTC Communications, and TPC of INFOCOM, GBLOBECOM, etc. His research interests include PHY/MAC designs, security, and mobile computing in cyber-physical systems.
\end{IEEEbiography}

\begin{IEEEbiography}{Yuanwei Liu (S'13-M'16)} (yuanwei.liu@qmul.ac.uk) is a Lecturer (Assistant Professor) in School of Electronic Engineering and Computer Science at Queen Mary University of London (QMUL) , London, U.K. He was a Postdoctoral Research Fellow at King's College London (KCL) , London, U.K. (Sep. 2016- Aug. 2017). He received the Ph.D. degree from QMUL in 2016. He received the M.S. and B.S. degrees from Beijing University of Posts and Telecommunications (BUPT)  in 2014 and 2011, respectively. He currently serves as an Editor of the IEEE Communications Letters and the IEEE Access.
\end{IEEEbiography}

\begin{IEEEbiography}{Zhiqing Luo} (zhiqing$\_$luo@hust.edu.cn) is currently pursuing his master's degree at school of electronics and information engineering, Huazhong University of Science and Technology, Hubei, China. Before that, he has received his Bachelor’s degree in communication engineering from Jilin University, Jilin, China, in June 2016. His research interests include PHY/MAC  designs and IoT security in wireless networks.
\end{IEEEbiography}

\begin{IEEEbiography}{Tao Jiang (M'06-SM'10)} (tao.jiang@ieee.org) is currently a Chair Professor with the School of Electronics Information and Communications, Huazhong University of Science and Technology, Wuhan, P. R. China. He has authored or co-authored more 300 technical papers in major journals and conferences and 5 books in the areas of wireless communications and networks. He was invited to serve as TPC Symposium Chair for the IEEE GLOBECOM 2013, IEEEE WCNC 2013 and ICCC 2013. He is served or serving as associate editor of some technical journals in communications, including in IEEE Transactions on Signal Processing, IEEE Communications Surveys and Tutorials, IEEE Transactions on Vehicular Technology, IEEE Internet of Things Journal, and he is the associate editor-in-chief of China Communications, etc.. He is a senior member of IEEE.
\end{IEEEbiography}

\begin{IEEEbiography}{Qian Zhang (M'00-SM'04-F'12)} (qianzh@cse.ust.hk) joined Hong Kong University of Science and Technology in Sept. 2005 where she is a full Professor in the Department of Computer Science and Engineering. Before that, she was in Microsoft Research Asia, Beijing, from July 1999, where she was the research manager of the Wireless and Networking Group. Dr. Zhang has published about 300 refereed papers in international leading journals and key conferences in the areas of wireless/Internet multimedia networking, wireless communications and networking, wireless sensor networks, and overlay networking. She is a Fellow of IEEE for ``contribution to the mobility and spectrum management of wireless networks and mobile communications". Dr. Zhang has received MIT TR100 (MIT Technology Review) worlds top young innovator award. She also received the Best Asia Pacific (AP) Young Researcher Award elected by IEEE Communication Society in year 2004. Her current research is on cognitive and cooperative networks, dynamic spectrum access and management, as well as wireless sensor networks. Dr. Zhang received the B.S., M.S., and Ph.D. degrees from Wuhan University, China, in 1994, 1996, and 1999, respectively, all in computer science.
\end{IEEEbiography}

\begin{IEEEbiography}{Arumugam Nallanathan (S’97–M’00–SM’05)} (arumugam.nallanathanz@kcl.ac.uk) was an Assistant Professor with the Department of Electrical and Computer Engineering, National University of Singapore, from 2000 to 2007. He served as the Head of Graduate Studies with the School of Natural and Mathematical Sciences, King’s College London, from 2011 to 2012. He is currently a Professor of Wireless Communications with the Department of Informatics, King’s College London, University of London. His research interests include 5G wireless networks, molecular communications, energy harvesting, and cognitive radio networks. He has authored nearly 300 technical papers in scientific journals and international conferences. He is an IEEE Distinguished Lecturer. He has been selected as a Thomson Reuters Highly Cited Researcher in 2016. Dr. Nallanathan received the IEEE Communications Society SPCE outstanding service award 2012 and the IEEE Communications Society RCC outstanding service award 2014. He served as the Chair of the Signal Processing and Communication Electronics Technical Committee of IEEE Communications Society. He is an Editor of the IEEE TRANSACTIONS ON COMMUNICATIONS and the IEEE TRANSACTIONS ON VEHICULAR TECHNOLOGY.
\end{IEEEbiography}

\bibliographystyle{IEEEtran}

\end{document}